\documentstyle[stwol]{article}

\def\Journal#1#2#3#4{{#1} {\bf #2}, #3 (#4)}

\def\NPB{{\em Nucl. Phys.} B}
\def\PLB{{\em Phys. Lett.}  B}

\def\ZPC{{\em Z. Phys.} C}

\def\be{\begin{equation}}
\def\ee{\end{equation}}
\def\bea{\begin{eqnarray}}
\def\eea{\end{eqnarray}}
\newcommand{\bi}{\bibitem}

\catcode`\@=11
\def\lsim{\mathrel{\mathpalette\@versim<}}
\def\gsim{\mathrel{\mathpalette\@versim>}}
\def\@versim#1#2{\vcenter{\offinterlineskip
\ialign{$\m@th#1\hfil##\hfil$\crcr#2\crcr\sim\crcr } }}
\catcode`\@=12

\begin{document}

\title{GAUGE-YUKAWA UNIFIED MODELS}

\author{ J. KUBO }

\address{Faculty of Natural Sciences, Kanazawa University, 920-11
  Kanazawa, Japan}

\author{ M. MONDRAGON }

\address{Depto. de F\'{\i}sica Te\'orica, Instituto de F\'{\i}sica,
  UNAM, Apdo. Postal 20-364, M\'exico 01000 D.F.}
 
\author{ G. ZOUPANOS }

\address{Physics Dept., Nat. Technical University, GR-157 80 Zografou,
  Athens, Greece}


\twocolumn[%
\begin{flushright}
  \begin{tabular}{l@{}}
    NTUA58/96\\
    hep-ph/9702391
  \end{tabular}
\end{flushright}
\vspace{1ex}  
\maketitle\abstracts{
Gauge-Yukawa Unification (GYU) is obtained in GUTs by searching for
renormalization  group invariant relations among gauge and Yukawa
couplings beyond the unification scale.  Of particular interest are
two supersymmetric GUTs, the finite and the minimal $SU(5)$ models.
Both models provided us,among others,with predictions of the top quark
mass which so far have passed successfully the tests of progressively more
accurate measurements.
}]

\section{Introduction}

An outstanding question of the theory of Elementary Particle Physics
is the plethora of free parameter of the Standard Model (SM).  The
traditional way of reducing the number of free parameters of the SM is
to require that the theory is more symmetric at higher scales.  This
approach has been applied, e.g. in GUTs, with a certain success. In fact the LEP
data \cite{abf} seem to suggest that we should require $N=1$
supersymmetry in addition to the minimal $SU(5)$ in order to obtain a
successful prediction for one of the low energy gauge couplings.

However, this attractive possibility has its limitations. As it is well
known increasing the gauge symmetry of a GUT (e.g. SO(10), $E_6$) does
not lead to a more  predictive theory for the low energy parameters.
This is due to the fact that an enlarged symmetry requires also
further breakings, which in general require additional free
parameters.  Alternatively we have suggested
\cite{mondragon1}$^-$\cite{kmtz2} that a natural gradual extension of the
GUT philosophy, in the prospect of increasing the predictability of
the low energy parameters of the theory, is to attempt to relate
the couplings of the gauge and Yukawa sectors, i.e. to achieve
Gauge-Yukawa Unification.  Searching for a symmetry that could provide
GYU one is naturally lead to consider $N=2$ supersymmetric
theories,\cite{fayet} which however proved to have more serious 
phenomenological problems than the SM.  The same criticism holds also
for superstring theories and composite models which could in principle
lead to relations among the gauge and Yukawa couplings.

\section{ Gauge-Yukawa Unification}

In our recent studies \cite{mondragon1}$^-$\cite{kmtz2} 
 we have considered
 the GYU which is based on the
principles of reduction of
couplings \cite{zimmermann1,kubo2,kmtz2}
 and in addition finiteness.\cite{mondragon1,finite1}$^-$\cite{finite3}
These principles, which are formulated in
perturbation  theory, are not explicit symmetry
principles, although they might imply symmetries.
 The former principle is based on the existence
of renormalization group
invariant (RGI) relations among couplings which preserve
 perturbative
renormalizability. Similarly, the latter one is based
on the fact that it is
possible to find RGI
 relations among couplings that
keep finiteness in perturbation theory,
even to all orders.\cite{finite3}
Applying these principles,
one can relate the gauge and Yukawa couplings,
thereby improving the
 predictive power of a model.
In what follows, we briefly outline the basic tool
of this GYU scheme and  its application to the most promising models.

A RGI relation among couplings can be expressed
in an implicit form
\be
\Phi (g_1,\cdots,g_N) ~=~0~,
\ee
which
has to satisfy the partial differential equation (PDE)
$$\mu\,d \Phi /d \mu ~=~
\sum_{i=1}^{N}
\,\beta_{i}\,\partial \Phi /\partial g_{i}~=~0,$$ 
where $\beta_i$ is the $\beta$-function of $g_i$.
There exist ($N-1$) independent  $\Phi$'s, and
finding the complete
set of these solutions is equivalent
to solve the so-called reduction equations \cite{zimmermann1},
\be
\beta_{g} \,\frac{d g_{i}}{d g} =\beta_{i}~,~i=1,\cdots,N~~,
\ee
where $g$ and $\beta_{g}$ are the primary
coupling and its $\beta$-function,
and $i$ does not include $g$.
Using all the $(N-1)\,\Phi$'s to impose RGI relations,
one can in principle
express all the couplings in terms of
a single coupling $g$.
 The complete reduction,
which formally preserve perturbative renormalizability,
 can be achieved by
demanding a 
 power series solution
\be
g_{i} = \sum_{n=0}\kappa_{i}^{(n)}\,g^{2n+1}~.
\label{powser}
\ee
The uniqueness of such a power series solution
can be investigated at the one-loop level \cite{zimmermann1}.
The completely reduced theory contains only one
independent coupling with the
corresponding $\beta$-function.
In supersymmetric Yang-Mills theories with
a simple gauge group, something more  drastic can happen;
the vanishing of the $\beta$-function
 to all orders in perturbation theory,
if all the one-loop anomalous dimensions of the matter fields
in the completely and uniquely reduced
theory vanish identically.\cite{finite3}

This possibility of coupling unification is
attractive, but  it can be too restrictive and hence
unrealistic. To overcome this problem, one  may use fewer $\Phi$'s
as RGI constraints.
This is the idea of partial
 reduction,\cite{zimmermann1,kubo2,kubo3} and the power series
solution (\ref{powser}) becomes in this case 
\bea
g_{i} &=& \sum_{n=0}\kappa_{i}^{(n)}(g_{a}/g)~ g^{2n+1}~,~\\
&&i=1,\cdots,N'~,~a=N'+1,\cdots,N~.\nonumber
\eea
The coefficient functions
 $\kappa_{i}^{(n)}$ are required to
be unique power series in $g_{a}/g$ so that the $g_{a}$'s can
be regarded as perturbations to the completely reduced
system in which the $g_{a}$'s identically vanish.
In the following,
 we would like to consider two very interesting models which are also
 representative of the two mentioned possibilities.\cite{mondragon1,kubo2}

\section { Gauge-Yukawa Unified Models}
\subsection { The $SU(5)$ Finite Unified Theory }

This is a $N=1$ supersymmetry Yang-Mills theory  based on
 $SU(5)$ \cite{finite2} which contains one ${\bf 24}$,
four pairs of (${\bf 5}+\overline{{\bf 5}}$)-Higgses and
 three
($\overline{{\bf 5}}+{\bf 10}$)'s
for three fermion generations. It has been done a complete reduction
 of the dimensionless parameters of the theory in favour of the gauge
 coupling $g$ and the unique power series solution \cite{mondragon1}
 corresponds to 
the Yukawa matrices
without intergenerational mixing, and yields
in the one-loop approximation
\bea
g_{t}^{2} =g_{c}^{2}~=~g_{u}^{2}&=&\frac{8}{5} g^2~,~\\
g_{b}^{2} ~=~g_{s}^{2} ~=~g_{d}^{2} &=& \frac{6}{5} g^2~,\\
g_{\tau}^{2} ~=~g_{\mu}^{2} ~=~g_{e}^{2}
{}&=&\frac{6}{5} g^2~,
\eea
where $g_i$'s stand  for the Yukawa couplings.
At first sight, this GYU seems to lead
to unacceptable predictions of the fermion masses.
But this is not the case, because each generation has
an own pair of ($\overline{{\bf 5}}+{\bf 5}$)-Higgses
so that one may assume \cite{finite2,mondragon1} that
after the diagonalization
of the Higgs fields  the effective theory is
exactly MSSM, where the pair of
its Higgs supermultiplets mainly stems from the
(${\bf 5}+\overline{{\bf 5}} $) which
couples to the third fermion generation.
(The Yukawa couplings of the first two generations
can be regarded as free parameters.)
The predictions of $m_t$ and $m_b$ for various $m_{\rm SUSY}$
are given in Table 1.

\begin{table*}
\begin{center}\caption{The predictions of the FUT
 $SU(5)$}\label{table-fut}
\vspace{0.4cm}
\begin{tabular}{|c|c|c|c|c|c|}
\hline
$m_{\rm SUSY}$ [GeV]   &$\alpha_{3}(M_Z)$ &
$\tan \beta$  &  $M_{\rm GUT}$ [GeV]
 & $m_{b} $ [GeV]& $m_{t}$ [GeV]
\\ \hline
$200$ & $0.123$  & $53.7$ & $2.25 \times 10^{16}$
 & $5.2$ & $184.0$ \\ \hline
$500$ & $0.118$  & $54.2$ & $1.45 \times 10^{16}$
 & $5.1$ & $184.4$ \\ \hline
\end{tabular}
\end{center}
\end{table*}

\subsection{ The minimal supersymmetric $SU(5)$ model}

The field content is minimal. Neglecting the
CKM mixing, one starts with six
Yukawa and two Higgs
couplings. We then require
GYU to occur among the Yukawa couplings of the third
generation and the gauge coupling.
We also require the theory to
be completely asymptotically free.
In the one-loop approximation, the GYU yields
$g_{t,b}^{2} ~=~\sum_{m,n=1}^{\infty}
\kappa^{(m,n)}_{t,b}~h^m\,f^n~g^2$ 
($h$ and $f$ are related
to the Higgs couplings). Where 
$h$ is allowed to vary from
$0$ to $15/7$, while $f$ may vary from $0$ to a maximum which depends
on $h$ and vanishes at $h=15/7$.
As a result, we obtain\cite{kubo2}
\bea
0.97\,g^2 \lsim &g_{t}^{2}& \lsim 1.37\,g^2~,\nonumber\\
{}~0.57\,g^2 \lsim g_{b}^{2}&=&g_{\tau}^{2}  \lsim 0.97\,g^2~.
\eea

We found \cite{kmoz,kmoz2} that consistency with proton decay requires
$g_t^2,~g_b^2$ to be very close to the left hand side values in the
inequalities.  In Table 2 we give the predictions for representative
values of $m_{SUSY}$.

\begin{table*}
\begin{center}\caption{The predictions of the minimal SUSY
    $SU(5)$}\label{table-afut} 
\vspace{0.4 cm}
\begin{tabular}{|c|c|c|c|c|c|c|c|}
\hline
$m_{\rm SUSY}$ [GeV]
& $g_{t}^{2}/g^2$ & $g_{b}^{2}/g^2$&
$\alpha_{3}(M_Z)$ &
$\tan \beta$  &  $M_{\rm GUT}$ [GeV]
 & $m_{b} $ [GeV]& $m_{t}$ [GeV]
\\ \hline
$300$ & $0.97$& $0.57$ & $0.120$  & $47.7$ & $1.8\times10^{16}$
  & $5.4$  & $179.7 $  \\ \hline
$500$ & $0.97$& $0.57$ & $0.118$  & $47.7$ & $1.39\times10^{16}$
  & $5.3$  & $178.9$  \\ \hline
\end{tabular}
\end{center}
\end{table*}

In all of the analyses above,
 we have used the
RG technique and regarded the GYU relations
 the boundary conditions  holding
at the unification scale $M_{\rm GUT}$.
We have assumed that
it is possible to arrange
the susy mass parameters along with
the soft breaking terms in such a way that
the desired symmetry breaking pattern
 really occurs, all the superpartners are
unobservable at present energies,
there is no contradiction with proton decay,
and so forth.
To simplify our numerical analysis  we have also assumed a
unique  threshold $m_{\rm SUSY}$ for all
the  superpartners.

Using the updated experimental data on the SM parameters, we have
re-examined the $m_t$ prediction of the two GYU $SU(5)$ models
described above.\cite{kmoz2} ~They predict

\noindent FUT:
\be
m_t = (183 + \delta ^{MSSM}_{m_t} \pm 5)~ GeV
\ee

\noindent Min. SUSY $SU(5)$:
\be
m_t = (181 + \delta ^{MSSM}_{m_t} \pm 3)~ GeV 
\ee
where $\delta ^{MSSM}_{m_t}$ stands for the MSSM threshold
corrections.  One obtains an idea about the magnitude of the
correction by considering the case that all superpartners have the
same mass $m_{SUSY}$ and $m_{SUSY} \gg \mu _H$, where $\mu_H$ describes
the mixing of the two Higgs doublets in the superpotential.  In that
case we found \cite{kmoz2} $\delta ^{MSSM}_{m_t} \sim - 1\%$.

\section{Discussion and Conclusions}

As a natural extension of the unification of gauge couplings provided by 
all GUTs and the unification of Yukawa couplings, we
have introduced the idea of Gauge-Yukawa
Unification. GYU is a functional relationship among the gauge and
Yukawa couplings provided by some principle.  In our studies GYU has
been achieved by applying the principles of reduction of couplings
and finiteness. 
The consequence of GYU is that 
in the lowest order in perturbation theory
 the gauge and Yukawa couplings above  $M_{\rm GUT}$
are related  in the form
\be
g_i  = \kappa_i \,g_{\rm GUT}~,~i=1,2,3,e,\cdots,\tau,b,t~,
\label{bdry}
\ee 
where $g_i~(i=1,\cdots,t)$ stand for the gauge 
and Yukawa couplings, $g_{\rm GUT}$ is the unified coupling,
and
we have neglected  the Cabibbo-Kobayashi-Maskawa mixing 
of the quarks.
 So, Eq.~(\ref{bdry}) exhibits a boundary condition on the 
the renormalization group evolution for the effective theory
below $M_{\rm GUT}$, which we have assumed to 
be the MSSM. 
We have shown \cite{kmoz,kmoz2} that it is
possible to construct some 
supersymmetric GUTs with GYU in the 
third generation that can
 predict the bottom and top
quark masses in accordance with the recent experimental data.
This means that the top-bottom hierarchy 
could be
explained in these models,
 in a similar way as 
the hierarchy of the gauge couplings of the SM
can be explained if one assumes  the existence of a unifying
gauge symmetry at $M_{\rm GUT}$. 

It is clear that the GYU scenario  is the most predictive scheme as far as
the mass of the top quark is concerned.
It may be worth recalling the predictions for $m_t$
of ordinary GUTs, in particular of supersymmetric $SU(5)$ and
$SO(10)$.  The MSSM with $SU(5)$ Yukawa boundary unification allows
$m_t$ to be anywhere in the interval between 100-200 GeV 
for varying $\tan \beta$, which is now a free parameter.  Similarly,
the MSSM with $SO(10)$ Yukawa 
boundary conditions, {\em i.e.} $t-b-\tau$ Yukawa Unification gives
$m_t$ in the interval 160-200 GeV. 
We have analyzed \cite{kmoz2} the infrared quasi-fixed-point behaviour of
the $m_t$ prediction in some detail. In particular we have seen that
the {\em infrared value} for large $\tan \beta$ depends on  $\tan
\beta$  and its lowest value is $\sim 188$ GeV.  Comparing this with
the experimental value $m_t = (176.8 \pm 6.5)$ GeV we may conclude
that the present data on $m_t$ cannot be explained from the infrared
quasi-fixed-point behaviour alone.

Clearly, to exclude or verify different GYU models,
 the experimental as well as theoretical uncertainties
have to be further reduced.
One of the largest theoretical uncertainties 
 in FUT  results
from the not-yet-calculated threshold effects 
of the superheavy particles.
Since the structure of  the superheavy 
particles is basically fixed,
 it will be possible to
bring these threshold effects under control,
which will  reduce the uncertainty of 
the $m_t$ prediction.
 We have been regarding $\delta^{\rm MSSM} m_t$ 
as unknown because we do not have 
sufficient information on the superpartner spectra.
Recently, however, we have demonstrated \cite{kmz-pert} how to extend
 the principle of reduction of couplings in a way as to include the
 dimensionfull parameteres.  As a result, it is in principle possible
 to predict the superpartner spectra as well as the rest of the
 massive parameters of a theory. 

\section*{Acknowledgments}

Presented by G. Zoupanos.  Partially supported by the E.C.
projects CHRX-CT93-0319, and ERBFMRXCT960090, the Greek project
PENED/1170, and the UNAM Papiit project IN110296.

\end{document}